\documentclass[sigconf,screen,nonacm,authorversion]{acmart}

\usepackage{graphicx}
\usepackage{multirow}
\usepackage{booktabs}
\usepackage[htt]{hyphenat}  

\usepackage{listings}
\usepackage{xcolor}
\lstdefinelanguage{json}{
    basicstyle=\ttfamily\scriptsize,
    numbers=left,                    
    numberstyle=\tiny\color{gray},    
    stepnumber=1,                     
    showstringspaces=false,
    breaklines=true,
    frame=single,
    literate=
     *{0}{{{\color{blue}0}}}{1}
      {1}{{{\color{blue}1}}}{1}
      {2}{{{\color{blue}2}}}{1}
      {3}{{{\color{blue}3}}}{1}
      {4}{{{\color{blue}4}}}{1}
      {5}{{{\color{blue}5}}}{1}
      {6}{{{\color{blue}6}}}{1}
      {7}{{{\color{blue}7}}}{1}
      {8}{{{\color{blue}8}}}{1}
      {9}{{{\color{blue}9}}}{1}
      {:}{{{\color{red}:}}}{1}
      {,}{{{\color{red},}}}{1}
      {\{}{{{\color{purple}\{}}}{1}
      {\}}{{{\color{purple}\}}}}{1}
      {[}{{{\color{purple}[}}}{1}
      {]}{{{\color{purple}]}}}{1},
}

\newcommand{\CloudHeatMap}{\textsc{CloudHeatMap}}

\begin{document}

\title{CloudHeatMap: Heatmap-Based Monitoring for Large-Scale Cloud Systems}

\settopmatter{authorsperrow=4}
\author{Sarah Sohana}
\email{sarahsohana@gmail.com}
\affiliation{%
  \institution{Toronto Metropolitan U.}
  \city{Toronto}
  \country{Canada}
}
\author{William Pourmajidi}
\email{william.pourmajidi@gmail.com}
\affiliation{%
  \institution{Toronto Metropolitan U.}
  \city{Toronto}
  \country{Canada}
}

\author{John Steinbacher}
\email{jstein@ca.ibm.com}
\affiliation{%
  \institution{IBM Canada Lab}
  \city{Toronto}
  \country{Canada}
}

\author{Andriy Miranskyy}
\email{avm@torontomu.ca}
\affiliation{%
  \institution{Toronto Metropolitan U.}
  \city{Toronto}
  \country{Canada}
}

\begin{abstract}
Cloud computing is essential for modern enterprises, requiring robust tools to monitor and manage Large-Scale Cloud Systems (LCS). Traditional monitoring tools often miss critical insights due to the complexity and volume of LCS telemetry data. This paper presents \CloudHeatMap, a novel heatmap-based visualization tool for near-real-time monitoring of LCS health. It offers intuitive visualizations of key metrics, such as call volumes and response times, enabling operators to quickly identify performance issues. A case study on the IBM Cloud Console demonstrates the tool's effectiveness in enhancing operational monitoring and decision-making. 
\end{abstract}

\maketitle

\section{Introduction}
Cloud computing has become a critical infrastructure for businesses, providing scalability and cost reduction. However, as Cloud adoption grows, monitoring these systems becomes increasingly complex~\cite{pourmajidi2017challenges,pourmajidi2019dogfooding,pourmajidi2021challenging}. Large-Scale Cloud Systems (LCS) comprise numerous interconnected microservices distributed across data centers (see~\cite{islam2021anomaly,islam2024anomaly} for an example), generating vast amounts of telemetry data~\cite{miranskyy2016operational} essential for understanding system health and performance. The sheer volume and complexity of these data challenge traditional monitoring tools, which often fail to provide timely, actionable insights~\cite{pourmajidi2017challenges,pourmajidi2019dogfooding,pourmajidi2021challenging}. This lack can lead to undetected performance issues, inefficient resource allocation, and, ultimately, system failures that impact service delivery~\cite{pourmajidi2017challenges,pourmajidi2019dogfooding,pourmajidi2021challenging}.

Traditional Cloud monitoring tools (e.g., DataDog, Dynatrace) offer limited root cause visibility and are often reactive, alerting after issues impact the system. In dynamic Cloud environments, delayed responses can cause significant downtime and resource inefficiency. According to Google's Site Reliability Engineering (SRE) principles, the four Golden Rules of Monitoring~---~traffic, errors, latency, and saturation~---~are essential metrics for assessing LCS health~\cite{4goldenrules}. However, most monitoring tools focus on the interface-level analysis of system components~\cite{blackbox}, detecting issues but not providing deeper insights  or predicting problems. While some advanced monitoring systems offer visual dashboards~\cite{microservice_monitoring}, they often lack support for exploratory health analysis. Operators need to delve into data patterns and understand component behaviours under varying conditions~\cite{baciu2017cognitive}. In today’s dynamic Cloud environments, early and proactive detection are critical~\cite{islam2020anomaly,islam2021anomaly, hrusto2022optimization,hrusto2023towards,hrusto2024autonomous,zhong2024detecting}; reactive approaches are insufficient. Thus, advanced monitoring solutions are needed that provide real-time insights and enable exploratory analysis to help Cloud Operators (Ops) teams identify issues before they escalate~\cite{pourmajidi2017challenges,pourmajidi2019dogfooding,pourmajidi2021challenging}.

The authors, drawing from their experience designing and using AIOps tools~\cite{pourmajidi2017challenges,pourmajidi2019dogfooding,pourmajidi2021challenging,islam2020anomaly,islam2021anomaly,islam2024anomaly}, can attest that while these tools are useful, they have limitations in root cause analysis and detection of persistent issues. Long-standing abnormalities can escape automated detection, requiring human operators to assess the broader system and make strategic decisions beyond local failures.

The need arises for a mechanism that allows for a complete, bird’s eye view of a complex system~---~one that human cognition can easily process. The proposed tool addresses these gaps by providing near-real-time, intuitive visualizations for deeper insights into system health, enabling operators to diagnose and prevent issues before they affect system reliability.  We are guided by the following research questions (RQs):

\textbf{RQ1}: How can we design visualizations that effectively support LCS monitoring and provide actionable insights?

\textbf{RQ2}: How do components across data centers respond to varying workloads, and how can we visualize these responses?

\textbf{RQ3}: How can we detect and visualize component issues in near-real-time for proactive management?

\textbf{RQ4}: How does our visualization tool impact LCS maintenance and operations?

We present \CloudHeatMap\footnote{\textbf{The tool is available via~\cite{CloudHeatMapSrc} and includes $\approx 24$ hours of synthetic data. A demonstration is available at~\cite{CloudHeatMapDemo}.}}, a novel heatmap-based\footnote{A grid heatmap maps data points as colours in two dimensions, where colour intensity represents data magnitude~\cite{wilkinson2009history}, helping identify patterns and relationships in large datasets, allowing for quicker decision-making.} visualization tool leveraging microservices telemetry data to visualize key performance metrics, such as call volumes and response times, with filtering options for HTTP and custom response codes.

We drew inspiration from the classical annunciator panels~\cite{isa1811979}, that were commonly used in power plants and aircraft control boards. These panels feature a grid of lights or buttons, each representing a system parameter. When a parameter changes, the light or button updates its colour or state, enabling operators to monitor processes efficiently and respond to alerts.

As we empirically found, heatmaps are valuable because they condense large amounts of data from complex systems into a simple grid, where colour highlights anomalies, allowing operators to quickly and intuitively detect unusual behaviour.

\CloudHeatMap\ (described in Section~\ref{sec:description}) offers a near-real-time system health overview, enabling operators to quickly identify hot spots and potential issues. The tool helps Cloud Ops make informed decisions about resource allocation, configuration changes, and problem resolution, ultimately enhancing Cloud services' reliability and efficiency. It also supports temporal analysis, helping operators track system behaviour over time. 

To evaluate tools's effectiveness, we conducted a case study (given in Section~\ref{sec:evaluation}) on the IBM Cloud Console, an LCS with significant operational demands. Results show that \CloudHeatMap\  enhances Cloud Ops' ability to monitor and maintain system health, providing insights previously difficult to obtain with traditional methods. This research contributes to Cloud computing by offering a scalable and practical solution for monitoring LCS.

\section{Tool Description}\label{sec:description}

\begin{figure}[tb]
\centerline{\includegraphics[width=0.9\columnwidth]{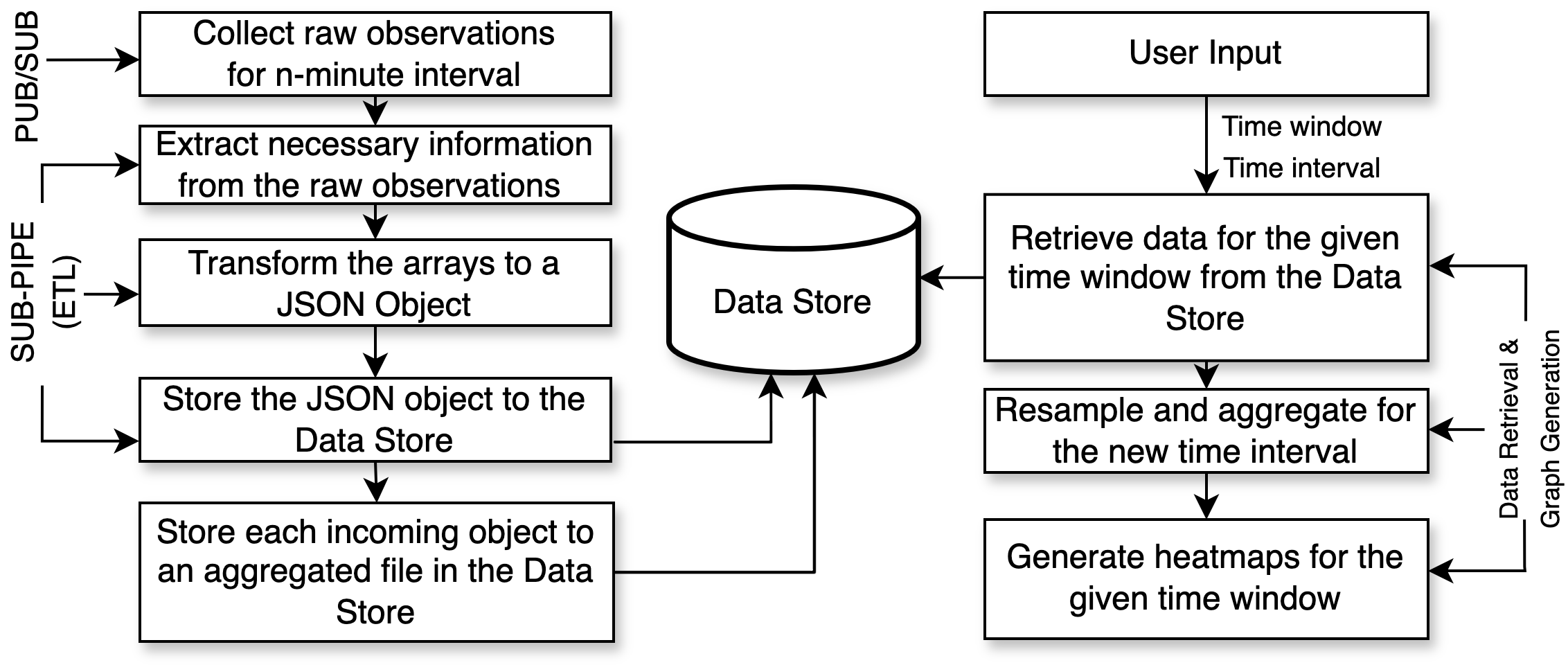}}
\caption{System diagram.}
\label{fig1}
\end{figure}

\CloudHeatMap\ user interface, described in Section~\ref{sec:visualization}, operates independently of the data collection method. Thus, for different application deployments, the telemetry collection process may vary. However, as long as the data are converted into the specified format~---~outlined in Section~\ref{sec:data_format}~---~the tool can visualize the data effectively. Figure~\ref{fig1} illustrates the sequence of actions involved, from data collection to heatmap generation. 

For completeness, Section~\ref{sec:original_sus_and_data_harvest} provides a brief overview of the data collection and processing mechanisms used in our studies. Note that the complexity of the collection process can vary based on the requirements and preferences of the organization responsible for maintaining and operating a given software.

\subsection{Original System and Data Harvesting}\label{sec:original_sus_and_data_harvest}
\subsubsection{System Analysis}\label{sec:analysis}

The first step  involved analyzing the LCS under study, the IBM Cloud Console,  a critical component of IBM's Cloud infrastructure~\cite{islam2021anomaly}, serving as the web front-end and orchestrator for the IBM Cloud platform~\cite{ibm}. Currently, the system consists of around 125 microservices deployed worldwide.  Each microservice generates millions of log records daily, encompassing various operational metrics, including response codes, call volumes, and response times~\cite{islam2021anomaly}.

Understanding the architecture and operational characteristics of the IBM Cloud Console was crucial for identifying appropriate metrics and effective visualization techniques. The complex distributed nature of the system and the volume of telemetry data posed significant challenges, making it ideal for testing the proposed visualization tool.

\subsubsection{Data Collection}\label{sec:collection}
Our data collection focuses on telemetry from the IBM Cloud Console, comprising metrics, logs, and traces emitted by microservices. Traces include essential request details such as timestamps, response codes, response times, and caller-callee microservice interactions\footnote{A user request triggers calls between microservices. The initiating microservice is the caller, and the receiving one is the callee.}. A robust, data-agnostic pipeline was designed using Redis Pub/Sub~\cite{birman1987exploiting} for handling high data volumes and IBM Cloud Object Storage (COS)~\cite{COS} for persistence.

Telemetry data, published at 1-minute intervals, was captured in Zipkin format~\cite{zipkin} (with support for extending to formats such as OpenTrace~\cite{opentrace} and OpenTelemetry~\cite{OpenTelemetry}). An extract, transform, and load (ETL) module extracted key fields (e.g., trace IDs, HTTP status codes) and generated summary statistics for the metrics in Table~\ref{tab1}.

The processed data was stored as nested JSON with indexed timestamps (see Section~\ref{sec:data_format} for format details) for efficient time series analysis. This structure enabled the creation of system health visualizations, including heatmaps.

\begin{table}[tb]
\caption{Metrics generated from the extracted data.}
\begin{center}
\resizebox{\columnwidth}{!}{%
\begin{tabular}{@{}l|l@{}}
\toprule
\textbf{\textit{Extracted Fields}} & \textbf{\textit{Metrics}} \\
\midrule
Count of Distinct Services & Call Volumes
\\
\midrule
\multirow{4}{4em}{Duration} & Average Response Time\\
& Maximum Response Time\\
& Minimum Response Time\\
& Standard Deviation of Response Time\\
\midrule
Status/Response Code & Percentage and value of status/response codes\\
\bottomrule
\end{tabular}
}
\label{tab1}
\end{center}
\end{table}

\subsubsection{Data Processing}\label{sec:processing}

Collected data was aggregated into user-defined intervals (e.g., 24 hours) via IBM Cloud Functions\footnote{Replaced by IBM Cloud Code Engine~\cite{CodeEngine} at the time of writing.}~\cite{functions}, producing files capped at 1~MB. Summary statistics, including sums and averages, were computed for each microservice. The aggregated data were resampled at desired time intervals (e.g., 15 minutes, hourly) using aggregation functions (e.g., compound mean and standard deviation, see~\cite{sohana2022thesis} for details).

Finally, the processed data were stored in IBM COS, providing secure, scalable access. A Python-based retrieval module interfaced with the COS API to load JSON data within specified time windows, converting them into two-dimensional matrices for further analysis.

\subsection{ Visualization Design} \label{sec:visualization}
The core part of the tool involved designing and implementing the heatmap visualization tool using the Plotly Dash library~\cite{dash, plotly}.

\begin{figure*}[tb]
\centerline{\includegraphics[width=2\columnwidth]{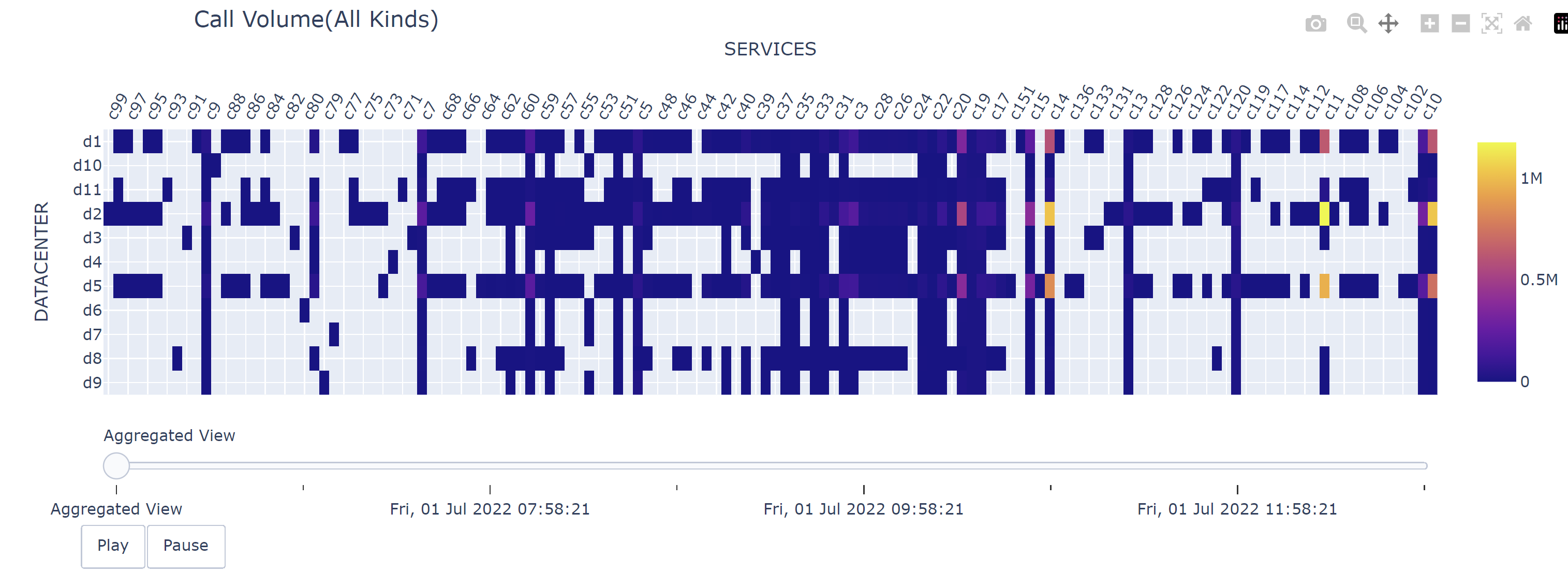}}
\caption{A heatmap comparing data centers and microservices, useful for assessing whether an issue has a greater impact on data centers.}
\label{fig2}
\end{figure*}

\begin{figure*}[tb]
\centerline{\includegraphics[width=2\columnwidth]{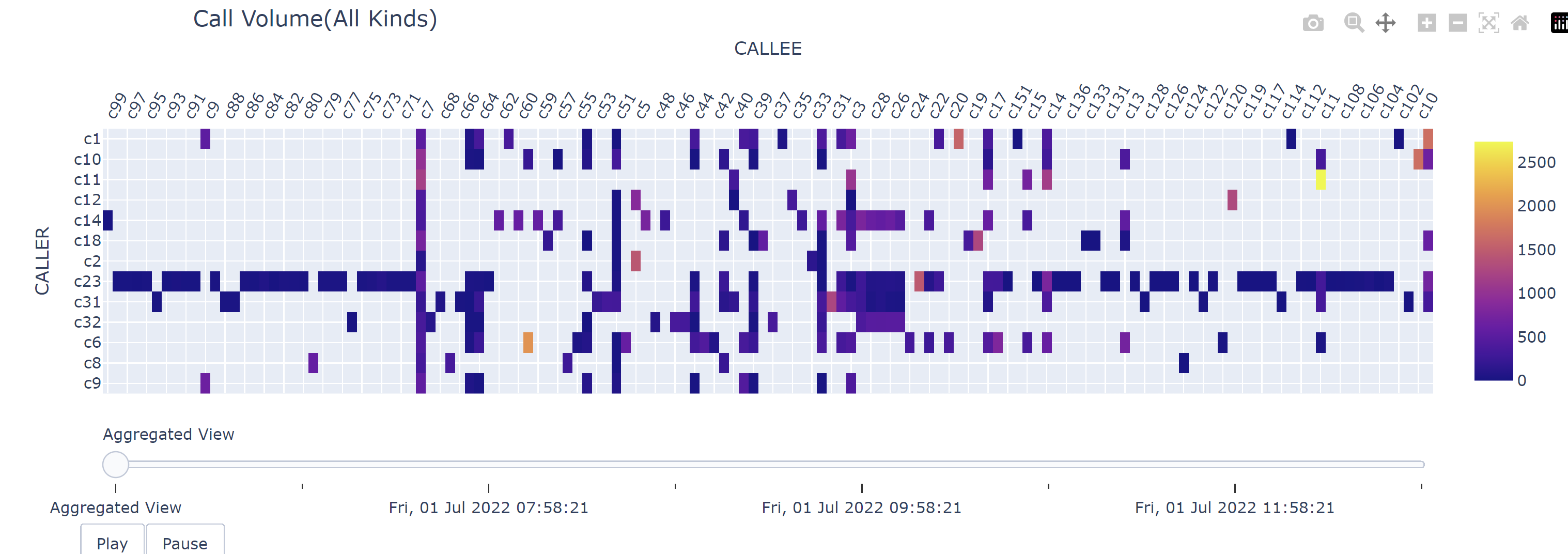}}
\caption{A heatmap showing caller-callee microservices, useful for identifying issues in the interactions between microservices.}
\label{fig3}
\end{figure*}

\CloudHeatMap\ visualizes call volumes and response times across the IBM Cloud Console's microservices and data centers. Figures~\ref{fig2} and~\ref{fig3} show sample heat maps using actual anonymised data. Figure~\ref{fig2} illustrates the relationship between data centers and services, with microservice names on the $x$-axis and data centers on the $y$-axis. Figure~\ref{fig3} depicts the caller-callee interactions, with callee microservices on the $x$-axis and caller microservices on the $y$-axis. Hovering with a mouse over a specific tile in the heatmap displays a window showing the numeric value of the tile along with its corresponding $x$ and $y$ coordinates, enhancing clarity.  Two key features are as follows.
\begin{enumerate}
    \item \textit{Interactive User Interface.} The tool offers a highly interactive UI with multiple filters, allowing customization based on graph type (data centers-services or caller-callee pairs), metrics (e.g., call volumes or average response time), and return codes. Users can switch between viewing absolute values or percentages and filter by value type and range.
    \item \textit{Temporal Analysis.} Users can play an animation displaying evolution of metrics over time, crucial for identifying trends and episodic issues that static views might miss. The first frame of the animation presents an aggregated view of the selected time period, offering a comprehensive snapshot, with subsequent frames showing behavioral changes.
\end{enumerate}

\CloudHeatMap\  is designed to highlight likely ``hot spots'' with areas of higher colour intensity on the heatmap (e.g., indicating higher call volumes, longer response times, or higher error rates). This helps operators quickly identify and address potential issues.

\subsection{Data formatting for visualization}\label{sec:data_format}
\CloudHeatMap\ consumes the data in the format shown in Listing~\ref{lst:json_structure}. For each timestamp (formatted as Unix timestamp in milliseconds), two dictionaries are recorded: \texttt{datacenter\_services} and \texttt{caller\_callee\_pairs}, corresponding to the data visualized in Figures~\ref{fig2} and~\ref{fig3}, respectively.

\subsubsection{Data for Figure~\ref{fig2}}
The \texttt{datacenter\_services} structure contains two primary identifiers: \texttt{app\_instance\_id} (mapped to data center in our case) and \texttt{microservice\_id}. 

The \texttt{microservice\_id} corresponds to individual microservices within the application. In Figure~\ref{fig2}, the $y$-axis represents the \texttt{app\_instance\_id}, while the $x$-axis represents the \texttt{microservice\_id}.

For each \texttt{app\_instance\_id} and \texttt{microservice\_id} pair, the data captures return codes. For each return code, we record a dictionary of statistics represented as pairs of the statistics name (e.g., \texttt{stats\_name\_1} or \texttt{min}) and a numeric value. These pairs are visualized as a heatmap in the figure.

\subsubsection{Data for Figure~\ref{fig3}}
The \texttt{caller\_callee\_pairs} structure is similar to \texttt{datacenter\_services}, but it captures relationships between pairs of microservices. Specifically, it records the interactions between a caller microservice (\texttt{caller\_microservice\_id}) and a callee microservice (\texttt{callee\_microservice\_id}), rendered on $y$-axis and $x$-axis of Figure~\ref{fig3}, respetively.

For each caller-callee pair, the data captures return codes and their associated statistics, which are then rendered as a heatmap to visualize the interactions.

\begin{lstlisting}[language=json, caption={JSON data structure injested by \CloudHeatMap.}, label={lst:json_structure}]
{"timestamp": [
  {
    "datacenter_services": {
      "app_instance_id": {
        "microservice_id": {
          "return_code_1": { 
            "stats_name_1": 1.0, "stats_name_2": 0.1},
          "return_code_2": { 
            "stats_name_1": 5.0, "stats_name_2": 6.2}
    }}},
      "caller_callee_pairs": {
        "caller_microservice_id_1": {
          "callee_microservice_id_1": {
            "return_code_1": {
              "stats_name_1": 1.1, "stats_name_2": 0.1}},
          "callee_microservice_id_2": {
            "return_code_1": { 
              "stats_name_1": 1.5, "stats_name_2": 0.5}}
  }}}]}
\end{lstlisting}

\subsubsection{Generalization Notes}\label{sec:generalization}
\textbf{The tool supports rendering telemetry from any application.} 
The \texttt{app\_instance\_id} identifies an application instance. The physical location of the application can vary based on the user's deployment strategy; however, \CloudHeatMap\ remains agnostic to these deployment specifics. \CloudHeatMap\ can monitor single or multiple instances of the same application, each distinguished by a unique \texttt{app\_instance\_id}.

Return codes can be any string-convertible format, not limited to HTTP status codes. If the application does not provide return codes, a placeholder return code can be hardcoded.

Statistical metrics can vary according to user preferences, provided that the statistic's name can be represented as a string and its value as a floating-point number.

The JSON files can be stored either in local storage (e.g., as demonstrated in our demo) or in cloud storage (e.g., as implemented in production)—depending on the user’s preference.

\section{Evaluation}\label{sec:evaluation}
We assessed \CloudHeatMap\ through a two-month case study on IBM Cloud Console (discussed in Section~\ref{sec:analysis}) in collaboration with the IBM Cloud Ops team, focusing on its ability to monitor and improve IBM Cloud Console's performance. For details of the evaluation of use cases, see~\cite{sohana2022thesis}.

\subsection{Rate Limiting Detection} Cloud microservices must remain within the rate limits to avoid service degradation. \CloudHeatMap\ revealed one microservice frequently exceeding its rate limit, with HTTP~429 ``Too Many Requests'' responses peaking at 40\%  across data centers in some time intervals. This visualization helped the Ops team quickly identify and address the issue.

\subsection{Detecting Components' Errors} 

Server-side errors (HTTP~5XX status codes) critically impact user experience. \CloudHeatMap\  enabled the Ops team to efficiently detect and address these errors. For example, it revealed services with a 100\% error rate, leading to immediate corrective actions, including removing obsolete services and rectifying active ones. The team prioritized services with significant call rates and planned to address lower-impact ones subsequently.

Not all endpoints return standard HTTP status codes. Some services use non-standard codes, like -1, for custom protocol interactions. \CloudHeatMap\ successfully visualized both HTTP and non-HTTP traffic, providing a comprehensive system view.

\subsection{Detecting Performance Degradation} 

User satisfaction is closely tied to interface responsiveness, with response times over two seconds degrading  experience~\cite{responsetime}. \CloudHeatMap\  identified (using caller-callee view) microservices contributing to latency issues, such as ``Dashboard-broker'' and ``Preferences'' interacting with a ``Cloudant'' database instance. Early detection allowed the Ops team to intervene before issues escalated into a critical failure. 

\subsection{Re-architecting Hot-spots}

Analysis of the ``Datalayer'' microservice using \CloudHeatMap\  identified inefficient calls, leading to improved caching strategies. The team re-architected the system, replacing the caching mechanism with an enhanced ``GraphQL'' layer, reducing call volumes and increasing flexibility for the front-end team.

\subsection{Cost Savings}

\CloudHeatMap\ contributed to cost-saving initiatives by identifying infrequently used microservices with low traffic, deployable on smaller, more cost-effective Kubernetes clusters, allowing the IBM Ops team to explore operational cost reductions through optimized resource deployment.

\section{Discussion}
\CloudHeatMap\ is a simple yet effective tool that equipped the IBM Ops team with actionable insights that traditional monitoring tools could not, revealing hidden issues and distinguishing between persistent and episodic problems. This proactive monitoring approach provides a more comprehensive view of system health, unlike the retrospective analyses of other platforms like Datadog and Dynatrace. Additionally, CloudHeatMap bridges the gaps left by machine-learning-based tools, which, despite their sophistication, often struggle with exploratory analysis~\cite{milo2020automating}.  By allowing operators and designers to reveal complex patterns requiring intervention, it enhances system comprehension and situational awareness.

These use cases align with some of Musa’s Operational Profiles analysis~\cite{musa93}, where observing client usage helps identify problematic components for testing and re-architecting, especially those frequently accessed or with high response times.

\CloudHeatMap\ answered RQ1 by providing visualizations that allowed operators to monitor three of the four golden signals recommended by Google SRE~---~traffic, errors, and latency. These visualizations enabled the Cloud Ops team to quickly identify and address performance issues, proving the visualizations to be actionable and insightful. For RQ2, the tool analyzed call volumes and workload patterns across different data centers, identifying bottlenecks and components struggling under varying workloads. This ensured operators could better understand components behaviours under dynamic conditions. For RQ3, \CloudHeatMap\  allowed near-real-time issue detection by enabling operators to monitor critical metrics as they evolved. The heatmap facilitated early identification of performance degradation, error spikes, and anomalies, allowing proactive issue resolution. For RQ4, the tool's insights directly impacted the maintenance process. The visualized data guided decisions like resource scaling and re-architecting high-traffic or error-prone components, improving system reliability and efficiency. These insights were validated through real-world use cases with the IBM Ops team.

\section{Threats to Validity}
The primary validity concerns, classified as per~\cite{yin2009case,wohlin2012experimentation}, are:

\textit{Construct Validity.} The tool was iteratively developed with feedback from the IBM Ops team and tested in production for two months, ensuring alignment with evaluation goals and mitigating potential threats to construct validity.

\textit{Internal Validity.} Rigorous testing and verification of the data were employed to minimize experimental errors.

\textit{External Validity.} The monitoring tool is not limited to one software system (see Section~\ref{sec:generalization}). While not tested on other products, it is designed to be generic and applicable to any LCS.

\section{Conclusion}
We proposed \CloudHeatMap, a heatmap-based tool to monitor LCS health, using the IBM Cloud Console as a software under study. The tool represents an advance in Cloud monitoring by providing actionable insights in real-time, allowing operators to make informed decisions about scaling, re-architecting, and cost-saving measures. Its approach to visualizing both HTTP and non-HTTP interactions makes it adaptable to various Cloud environments. We believe \CloudHeatMap\  will interest practitioners and academics and contribute to monitoring complex large-scale software systems. 

Moving forward, we will extend the tool to the fourth Golden Rule of Monitoring~---~saturation~---~to provide a more complete picture of system health. Furthermore, aggregating caller-callee pairs into call chains for graph-based root cause analysis will further enhance the tool's diagnostic capabilities.

\balance
\bibliographystyle{ACM-Reference-Format}
\bibliography{references}

\end{document}